\documentclass[sigconf]{acmart}

\AtBeginDocument{%
  \providecommand\BibTeX{{%
    \normalfont B\kern-0.5em{\scshape i\kern-0.25em b}\kern-0.8em\TeX}}}

\setcopyright{acmcopyright}
\copyrightyear{2021}
\acmYear{2021}
\acmDOI{10.1145/3461702.3462599}

\acmConference[AIES '21] {Proceedings of the 2021 AAAI/ACM Conference on AI, Ethics, and Society}{May 19--21, 2021}{Virtual Event, USA.}
\acmBooktitle{Proceedings of the 2021 AAAI/ACM Conference on AI, Ethics, and Society (AIES '21), May 19--21, 2021, Virtual Event, USA}
\acmPrice{15.00}
\acmISBN{978-1-4503-8473-5/21/05}

\usepackage[inline]{enumitem}
\usepackage{balance}

\newcounter{takeawayCounter}
\newcommand{\takeaway}[1]{\addtocounter{takeawayCounter}{1}{\thetakeawayCounter}}
\newcommand\takeawaywidth{22}

\begin{document}
\fancyhead{}

\title{Becoming Good at AI for Good}

\author{Meghana Kshirsagar}
\authornote{Equal first author contribution.}
\affiliation{%
\institution{Microsoft AI for Good}
\country{USA}}

\author{Caleb Robinson}
\authornotemark[1]
\affiliation{%
\institution{Microsoft AI for Good}
\country{USA}}

\author{Siyu Yang}
\authornotemark[1]
\affiliation{%
\institution{Microsoft AI for Good}
\country{USA}}

\author{Shahrzad Gholami}
\affiliation{%
\institution{Microsoft AI for Good}
\country{USA}}

\author{Ivan Klyuzhin}
\affiliation{%
\institution{Microsoft AI for Good}
\country{USA}}

\author{Sumit Mukherjee}
\affiliation{%
\institution{Microsoft AI for Good}
\country{USA}}

\author{Md Nasir}
\affiliation{%
\institution{Microsoft AI for Good}
\country{USA}}

\author{Anthony Ortiz}
\affiliation{%
\institution{Microsoft AI for Good}
\country{USA}}

\author{Felipe Oviedo}
\affiliation{%
\institution{Microsoft AI for Good}
\country{USA}}

\author{Darren Tanner}
\affiliation{%
\institution{Microsoft AI for Good}
\country{USA}}

\author{Anusua Trivedi}
\affiliation{%
\institution{Microsoft AI for Good}
\country{USA}}

\author{Yixi Xu}
\affiliation{%
\institution{Microsoft AI for Good}
\country{USA}}

\author{Ming Zhong}
\affiliation{%
\institution{Microsoft AI for Good}
\country{USA}}

\author{Bistra Dilkina}
\affiliation{%
\institution{University of Southern California}
\country{USA}}

\author{Rahul Dodhia}
\affiliation{%
\institution{Microsoft AI for Good}
\country{USA}}

\author{Juan M. Lavista Ferres}
\affiliation{%
\institution{Microsoft AI for Good}
\country{USA}}

\renewcommand{\shortauthors}{Kshirsagar, Robinson, and Yang, et al.}

\begin{abstract}
AI for good (AI4G) projects involve developing and applying artificial intelligence (AI) based solutions to further goals in areas such as sustainability, health, humanitarian aid, and social justice. Developing and deploying such solutions must be done in collaboration with partners who are experts in the domain in question and who already have experience in making progress towards such goals. Based on our experiences, we detail the different aspects of this type of collaboration broken down into four high-level categories: communication, data, modeling, and impact, and distill eleven takeaways to guide such projects in the future. We briefly describe two case studies to illustrate how some of these takeaways were applied in practice during our past collaborations.
\end{abstract}



\maketitle

\section{Introduction}

Advances in artificial intelligence (AI) and computing power have given rise to powerful AI tools ubiquitous in many people's personal and professional lives. These abilities are integrated into our phones and computers, and are driven mainly by businesses that have productized advances in AI at massive scales. Many of these tools are broadly available and provide some social benefit (e.g., search engines, navigation tools). However, the promise of AI to improve lives and protect vulnerable people and ecosystems has not yet reached its potential. 

AI for Good (AI4G) is a movement within the larger field of AI that aims to develop and use AI methods to further progress towards goals in sustainability, health, humanitarian aid, and social justice, guided loosely by the UN Sustainable Development Goals (SDGs) and priorities within local communities. Excellent literature reviews on the topic are offered by~\cite{shi2020artificial,rolnick2019tackling,cowls2019designing,vinuesa2020role}. 
A key difference from commercial applications of AI is that those AI4G problems and successes are often not defined by market need, but rather by non-profits, social enterprises and governments seeking to solve problems that have not found solutions in the private sector. 
For example, researchers in the field of computational sustainability~\cite{gomes2019computational} develop and apply methods to tackle problems such as wildlife conservation~\cite{fang2019artificial}, bioacoustics~\cite{zhong2020multispecies}, bird-migration tracking~\cite{sheldon2013approximate} and poverty detection~\cite{jean2016combining}.
In recent years, a number of pieces of criticism have been directed at the AI4G movement~\cite{berendt2019ai,green2019good,moore2019ai}.
While these critiques raise important concerns such as the bias of models trained on limited data, shifting attention away from root causes of societal problems, and a paternalistic understanding of the affected community, the discussion largely centers on the difficulty of defining what is ``good'' in our societal context.

In this article we distill first-hand experiences from our research lab focused on AI4G projects spanning several application areas over two years. Cognizant of the complexity of problems in the AI4G domain and our expertise restricted to the technical side of AI (statistics, modeling, and engineering), we collaborate extensively with external \textit{partner organizations} (PO) to define good outcomes for our projects, source and curate data, and realize real-world impact from our modeling solutions. These \textit{AI4G projects} contribute to solving problems in two ways: we develop and apply AI techniques to accelerate previously manual tasks such as data processing to enable the PO to arrive at their solution faster, and we analyze and model collected data for additional insights. The collaborative and practical nature of such projects means that the \textit{deliverables} are not just model weights, source code, and technical papers; they crucially involve working with these POs (whose technical capabilities and infrastructure vary greatly) to develop workable engineering solutions for deployment that respect resource constraints that POs face, as well as communicating and documenting our solutions -- and their limitations -- for the domain experts outside of computer science and engineering who use our models to impact society.

We highlight challenges that are more pronounced in AI4G projects compared to machine learning (ML) projects in the academic and corporate spheres, outline strategies we have learned for undertaking such projects, and reflect on difficulties we have faced measuring our impact. We break these down into four sections in the rest of the discussion: communication, data, modeling, and impact. Finally, we describe two case studies, i.e. \textit{AI4G projects}, that exemplify these difficulties and how we approached them in a real-world setting. 

\section{Communication}

The relationship and interaction between data scientists and POs -- who are the domain experts that define problems, curate data, and act on model outputs -- is an important first topic. Domain experts sometimes have decades worth of experience working in a problem area. Communicating all of this accumulated knowledge to data scientists within a few days or weeks during project planning can be difficult, but data scientists must be willing and ready to incorporate this knowledge into their modeling approaches. While straightforward, accurate bi-directional communication at all stages of an AI4G project is essential for its success, we focus below on areas where data scientists may need to drive the conversation with the PO.

\subsection{Setting realistic expectations from AI}

It is often the case that POs have inflated expectations about the capabilities of modern AI-based techniques due to the hype surrounding the field\footnote{For example, a Gartner 2020 report on emerging technologies places AI at the peak in terms of inflated expectations~\cite{gartner2020}.} and its misrepresentation in the media~\cite{marcus2019hype}. In our experience, when initially proposing and scoping projects, some POs may believe that there are pre-existing AI tools that can be immediately adapted for a niche purpose with little to no training data (cf.~\cite{chui2018ai}). However, our experience also shows that POs respond well to open, honest communication about AI's capabilities, or the need for (labeled) training data. Early conversations often involved directing POs' expectations away from a model that achieves all of their goals, which would require more training data than is currently available, to more targeted models for which appropriate and sufficient training data is available, and which will still bring them significantly closer to their goals. For certain use cases, an adequate and achievable approach is to use AI as a complementary tool to streamline and accelerate current workflows, rather than entirely supplanting them. 

One such example we have worked on involved a PO who initially approached us about building a natural language processing (NLP) model to extract very nuanced author intent from decontextualized 1- or 2-sentence texts in a small corpus of unlabeled documents. After gaining a better understanding of the specific use case for the model, we worked with the PO to reframe the problem in terms of multi-label topic classification, and also worked with them to label the data. The result was a model that allowed the PO to incorporate an entirely new data source (text) into a broader initiative focused on quantifying human activities' impacts on environmental resources. 

In addition to discussing the general limitations of AI-based methods with regards to what can be achieved, depending on the PO's domain and technical expertise, it may also be crucial to make them aware of more concrete issues encountered while building ML models. These include overfitting in small data regimes, model bias, generalization issues after deploying a model, adversarial attacks, data and model privacy concerns, limitations of interpretable models, etc. For example, lesion shape irregularity is one of the most critical features in the clinical diagnosis of melanoma~\cite{abbasi2004early}. On the other hand, recent studies have determined that convolutional neural networks are negatively-biased in capturing shape-related information from images~\cite{Baker2020LocalFA,geirhos2019imagenettrained}. Clinical researchers who may wish to develop a convolutional model for melanoma detection are likely to be unaware of this finding, and effective communication of this knowledge may facilitate the development of alternative approaches. We believe that knowledge transfer should be a core component of an AI4G project. 

In contrast to these positive cases, there may be circumstances in which POs need to be informed when their goals -- even with proper reframing -- may not be achievable with AI.
For example, a model trained to detect fish species from underwater cameras may be highly accurate in identifying a few common species, but the researchers may like to detect very rare species if, for example, detection of the rare species is key to deciding whether some economic development opportunity is allowed. In these cases, model creation may need to be postponed for time-consuming data collection and labeling. And in some cases it may be necessary to ask if the current state of machine learning is able to meaningfully help the PO in meeting its goals.

\begin{center}
\fbox{
\begin{minipage}{\takeawaywidth em}
\textbf{Takeaway \textit{\takeaway}.} Educating POs about AI's limits and opportunities is a core part of an AI4G project. Potentially unrealistic expectations for AI can often be reframed into achievable goals that streamline the PO's workflows.
\end{minipage}}
\end{center}

\subsection{Project scoping and implementation} \label{sec:communication:scoping}

Collaborations between POs and technical experts on AI4G projects are typically short-term or phased, hence it is important to set priorities and expectations before beginning a project. For example, one project we have worked on involves counting herd animals (e.g. cattle) over large areas from satellite imagery to monitor the effectiveness of conservation policies. This is an expensive task to perform manually, but can potentially be automated with computer vision models. Instance segmentation of the animals is a straightforward approach, however it is risky due to the lack of labeled data or low spatial resolution of the imagery. Less straightforward modeling approaches, such as coarser animal density estimation models, can satisfy the same project goals, but may take more effort to define upfront. Such approaches may also be informed by domain knowledge -- herd animals move in groups, resulting in dense crowds more easily identifiable in imagery than individual animals. The back-and-forth communication with the PO to understand the goal and explore \textit{appropriate} modeling solutions is a crucial part of the project life cycle.

We summarize questions to explore when scoping a project with a PO as a guideline for future projects:

\begin{enumerate}

\item Setting accurate goals: is the project oriented towards the prediction or estimation of quantities, or is it towards the visualization and representation of data? If the project is oriented towards prediction/estimation, then \textit{what} will be done with the output from the model, \textit{who} will use it (say, for decision-making), and \textit{where} will it be deployed? If the project is oriented towards visualization/representation or to interpretability/inference, \textit{what} format is expected, \textit{who} will be consuming the end result, and -- again -- \textit{where} will it be deployed?

\item What are the resource constraints of the PO? The compute and storage resources available and the type of device where the solution will be deployed (cloud-based, battery-powered devices) should be taken into account to determine what modeling solutions are possible.

\item What is the technical expertise available at the PO? This will determine the amount of support available for maintaining a deliverable beyond the duration of the project.

\item Does the problem require developing novel machine learning techniques? This will inform the project timeline and the risk. It could present an opportunity to identify larger research challenges and bring them into the ML community at large.

\item To what degree is data/model privacy a concern? The data available to the PO can often contain personally identifiable information about individuals. This would not only make the dataset not releasable for the sake of reproducibility, but also limit the public release of models since they may be subject to privacy attacks~\cite{shokri2017membership}.

\item Is precise information on the data well documented? The PO should convey the assumptions and the procedures underlying data collection. If the data has already been processed, the processing steps should be communicated because certain pre-processing steps can introduce bias in the data~\cite{garcia2015data}.

\item How could the PO's domain knowledge be incorporated in modeling? During the project development stage, it may be necessary to not rely entirely on data-driven techniques (for instance, for feature learning), but also utilize PO's knowledge of domain-relevant metrics and features that have been identified as significant in their field. For example, in radiology, dozens of hand-crafted radiomic features have been previously found to be significant predictors of cancer survival and response to therapy~\cite{Avanzo2017BeyondIma}.
\end{enumerate}

\begin{center}
\fbox{
\begin{minipage}{\takeawaywidth em}
\textbf{Takeaway \textit{\takeaway}.} To ensure we develop solutions that are practically useful, project scoping needs to be an ongoing dialogue with the PO.
\end{minipage}}
\end{center}
\label{sec:communication}

\section{Data}

Most AI4G projects start with the PO sharing a dataset they have previously collected and labeled, or pointing out public datasets that are relevant to the problem. While specific funding opportunities for creating datasets for ML applications and novel methods to take advantage of weakly labeled public data~\cite{wang2020weakly, zhou2018brief} or generated data~\cite{beery2020synthetic} exist, exploiting these is often not possible. This is in contrast to enterprise applications where ML teams are also responsible for collecting training data and have budget allocated for curating datasets with specific ML tasks in mind. In this section we reflect on recurring challenges related to data availability and quality in the context of AI4G projects; for an in-depth discussion of data issues present in high-stakes AI applications, refer to \cite{sambasivan2021everyone}.

\subsection{Adapting to previously collected datasets}

Limitations in the data collection process often influence the degree to which an AI4G project can be successful, possibly more so than they affect the outcomes of commercial or academic AI projects, which have more control over data collection.

First, there is a discrepancy between the purpose of data collection by POs that have the goal of solving an application problem and data collection by groups that have the goal of creating a generalizable model. The POs need not necessarily care about the metadata associated with data points that they label for furthering their goals, while such metadata may be important for quantifying how a model trained on such data will generalize. Put differently, if data collection is conducted by a PO with a focus on the content, rather than technical specifications of the data, then this can cause problems in post-hoc modeling steps.

Extending the example of counting herd animals from the last section, the PO might label herd animals from satellite imagery at a variety of spatial resolutions. The imagery could be at a spatial resolution of 0.1, 0.3, or 0.5 m/pixel; as long as the PO can count the types of animals of interest over a given area and date, they can meet their goals. On the other hand, a machine learning model that has been trained to identify herd animals from only 0.1 m/pixel imagery will likely not generalize to 0.5 m/pixel imagery as objects will be 5 times smaller in each dimension. The metadata associated with the labeled data is necessary to inform modeling decisions (e.g. augmenting the scale of satellite data and labels during model training will allow the model to generalize over a range of scales). This is in contrast to settings where data is collected specifically for the purpose of building effective models, where metadata would be considered explicitly at the data collection stage.

In addition to the completeness of metadata, quality and consistency of data collection, which do not affect experts' ability to discern the content but which pose additional generalization challenges to machine learning models, are a frequent issue. For example, in a study applying speech feature modeling to analyze mental health issues, the recording of conversations between military personnel with suicide risk and their therapists is used to predict their emotional bond~\cite{nasir2017complexity}. Here, the consistent use of dedicated microphones for the two speakers in a controlled environment was found to improve modeling outcomes, however this aspect of data collection would not affect the manual analysis of the recordings.

Second, AI4G projects often involve sensitive data necessitating the adherence to strict ethical and legal restrictions that make using such data more difficult. For example, many agencies would like to train computer vision models to detect images of Child Sexual Abuse Material (CSAM), but databases such as the Child Abuse Image Database (CAID) maintained by the UK Government are not accessible to other organizations~\cite{Broadhurst2020}. 
Other examples include applications with healthcare data -- models that identify pulmonary features from chest x-ray imagery need to be trained with labeled chest x-rays, a data source that initially requires pairing patient records with their chest x-rays. These sensitive applications require specialized privacy-preserving modeling methods and a layer of complexity that other applications do not entail. POs may choose to share synthetic data generated using generative models instead of real data but these approaches also may be prone to membership inference attacks~\cite{hayes2019logan}. Recent work on privacy preserving ML has developed both defenses~\cite{mukherjee2019protecting,xie2018differentially} and ways to measure the privacy risks from releasing models~\cite{liu2020mace,jayaraman2020revisiting}.

Third, the amount and quality of data available in some AI4G projects will be limited. This is not specific to AI4G projects, however is worth mentioning due to the frequency with which such projects come up. For example, projects that involve detecting poultry barns, solar farms, or other relatively uncommon features from satellite imagery require having a labeled dataset of such features. However, creating labeled data in these applications is expensive as it requires annotators to first find examples of the objects in question over large landscapes before labeling them appropriately. In our experience, these type of satellite image annotations will not be in a format that is immediately useable (e.g. point labels for a segmentation problem), or will be a biased sample (e.g. many labels from a specific area rather than a sample of labels from a broad area).

Finally, open datasets often have significant data curation issues. 
For example, a Kaggle dataset for predicting the outcome of pregnancies in India has been shown to be missing key data from the original survey, resulting in misleading predictions~\cite{trivedi2019risks}. POs that want to help reduce infant mortality can be misled by such datasets or the promise of effective models from open competitions that use such data. Another issue is the lack of query infrastructure around public datasets which creates significant friction in projects that might use such data. A positive example is Google Earth Engine~\cite{gorelick2017google}, which allows researchers to quickly query across its collection of public satellite imagery, visualize sample patches, and assess if the data is of sufficient quantity and resolution for the intended analysis.

\begin{center}
\fbox{
\begin{minipage}{\takeawaywidth em}
\textbf{Takeaway \textit{\takeaway}.} Datasets in AI4G projects may not be immediately useful for creating models. When creating models with such data, it is important to understand the associated metadata, collection process, and any security or privacy concerns.
\end{minipage}}
\end{center}

\subsection{Dealing with subjective data annotation}

The variables of interest in several socially important domains involving human perception and decision-making are ambiguous and ill-defined.
Different annotators might interpret the definition of labels differently, leading to inconsistent and noisy labels.
For example, the colloquial meaning of ``depression'' might be different from its meaning in a clinical context~\cite{yazdavar2017semi}.
The data annotation process for an AI system aimed at diagnosing clinical depression should be cognizant of the difference.
Creating a taxonomy of labels could minimize the annotator's subjectivity.
This also requires transparency in the interpretation of the labels and clearly communicating its limitations during different stages of the project life cycle, including annotation, modelling and deployment to minimize the semantic ambiguity of labels.
An in-depth discussion on ambiguous labels in the context of computational social sciences can be found in~\cite{chen2018using}.
When subjectivity of the labels are inherent due to the variability in human perception, it should be a standard practice to employ multiple annotators for each example and judge the feasibility of the task by evaluating inter-annotator agreement.
Several methods have been proposed to obtain estimates of the true ground truth labels from the noisy/subjective labels collected from multiple annotators~\cite{raykar2010learning, nasir2015redundancy}.

\begin{center}
\fbox{
\begin{minipage}{\takeawaywidth em}
\textbf{Takeaway \textit{\takeaway}.} In several socially important domains, labels suffer from subjective annotation. Such situation should be identified upfront to avoid introducing inconsistencies in the modeling pipeline.
\end{minipage}}
\end{center}

\subsection{Creating training and test sets with the application scenario in mind}

Poor choices of training, validation and test set splits can result in an estimated model performance that does not reflect actual performance when deployed (for other lessons learnt in evaluating model performance, see section \ref{sec:modeling-eval-and-metrics}). This is especially relevant in humanitarian aid and conservation applications where models are expected to generalize well spatially and/or temporally. 

For instance, with marine mammal sound detection~\cite{zhong2020beluga}, while generating train/test splits, consideration should be given to different types of underwater and anthropogenic noises such as those from commercial ship generators, mining, aircraft, and seasonal effects on ocean waves. As another example, the xBD dataset associated with the xView Challenge~\cite{Gupta_2019_CVPR_Workshops} is a large-scale public dataset designed to enable building damage assessment for humanitarian assistance and disaster recovery. It consists of data from 19 disasters from around the world between 2011 and 2019. However, scenes from the 19 disasters are present in all of the official train, test and hold-out splits, whereas it would be more useful to report performance on unseen locations as the next disaster will likely strike elsewhere. The same is true for the SpaceNet Challenge Series for building footprint extraction, where the default splits created by the challenge's utilities contain overlapping locations~\cite{van2018spacenet}. Subsequent studies have shown that performance drops drastically when applying a trained building damage classifier to an unseen location, even within the same region~\cite{valentijn2020multi, hao2020attention}. As another example requiring spatial generalization, until very recently, all studies of animal species classification on camera trap images were split across sequences of images but not across locations. This results in precision and recall metrics of greater than 90\%~\cite{norouzzadeh2018automatically}. In reality, performance is much worse if we split the data by camera location, and even worse if we split by ecosystem~\cite{beery2018recognition,schneider2020three}.

Many past studies in wildfire risk prediction, another problem important to both disaster relief and environmental conservation, assume certain random variables to be independent and identically distributed in the evaluation phase~\cite{safi2013prediction, rodrigues2014insight, castelli2015predicting}. However, natural hazards like wildfires are stochastic events with spatio-temporal dimensions, and evaluating models of such events based on randomized training and test splits leads to information leakage, misleading organizations who operationalize such models~\cite{gholami2021where}.

In applying machine learning to medical imaging, \citet{zech2018variable} found that a dataset of chest x-rays curated for screening pneumonia cases could be used to train a model to accurately predict which hospital system the x-ray comes from, indicating that the pneumonia detection model developed from the dataset could have been aided by features not related to the medical condition. The training and test splits should be chosen to measure how well the model will work for unseen x-ray machines.

\begin{center}
\fbox{
\begin{minipage}{\takeawaywidth em}
\textbf{Takeaway \textit{\takeaway}.} Carefully consider how to split the data into training and test sets so that the model's ability to generalize to unseen instances of input is measured.
\end{minipage}}
\end{center}
\label{sec:data}

\section{Modeling}

The models used in AI4G contexts usually involve a different set of requirements and constraints compared to general AI applications. First, AI4G models are developed in applied ML contexts. Most of the models are developed with domain-specific motivations and limitations in mind. In consequence, models developed for mainstream ML fields, such as NLP or computer vision, require cautious adaptation and deployment to the specific domain. Furthermore, the process of model development is motivated first by application requirements instead of pure novelty or state-of-the-art performance.

\subsection{Incorporating domain expertise}

Domain expertise from the PO can help in model development as POs often have decades of experience and accumulated knowledge in defining and solving related problems. Specifically, domain expertise is useful in: \begin{enumerate*}[label=\roman*)] \item determining adequate features and data representations, \item enforcing inductive bias and regularization in models, \item choosing simplified parameterizations, \item interpreting the learned models and outputs.\end{enumerate*}

Domain expertise is invaluable for collecting features that are relevant to a problem. This is especially relevant in AI4G problems where there are few samples compared to the number of features or where informative features are mixed with noisy features. For instance, to predict the late effects of chemotherapy on cancer patients, we found that including all chemotherapy drugs gave us a higher predictive performance as compared to prior work. However, clinical researchers know that only certain drugs have been clinically linked to certain late effects in other prior analyses. This suggests that the additional confounding features were probably spurious contributors to predictive performance and hence should not be included in the model.

In addition to helping determine data representation, domain expertise can help improve performance by embedding specific knowledge, often in the form of inductive bias or regularization, in model design. For example, finding promising solar cell technologies is often a difficult and a time-consuming task. A solar cell consists of a stack of various semiconductor materials, where each layer performs a certain electrical and optical function and fabrication parameters are optimized for maximizing solar energy captured. A machine learning model can avoid the need for resource-intensive physical experiments and accelerate the parameter optimization step. Combining a supervised machine learning model with a physical model of solar cell operation calibrated by an expert allowed for a model regularization method based on physical principles~\cite{ren2020embedding} and an order of magnitude reduction in the time and resources required to create a solar cell~\cite{ren2020embedding,oviedo2020bridging}.

Further, AI4G projects often involve collaborations spanning multiple countries and POs in a different country are likely to face unique challenges that only local experts would be aware of. For example, work on the anti-poaching PAWS project~\cite{fang2016deploying} uses the local knowledge of park rangers to constrain predicted search patterns to areas that can be feasibly visited.

If an interpretable model is used, then domain experts may be able to use aggregated model predictions to draw larger conclusions about a problem. For example, in our collaboration studying the food security in low-resource communities in Malawi based on survey panel data of households, domain experts were able to use the outputs of interpretable models to recognize spatial and seasonal patterns associated with the food security status of the communities and villages. These community-level insights can help local governments manage their resources more efficiently across the communities and over time.

\begin{center}
\fbox{
\begin{minipage}{\takeawaywidth em}
\textbf{Takeaway \textit{\takeaway}.} Endeavor to incorporate the PO's domain expertise in model development when possible through methods such as feature selection and engineering, model choice, and model regularization.
\end{minipage}}
\end{center}

\subsection{Model development with resource constraints}

Since deployed models are maintained by the PO who often has less resources than enterprises focused on mainstream ML applications, resource constraints once the model is operationalized limit the choice of models. In addition to the financial cost of running sophisticated models on potentially large datasets, deploying to remote regions in battery-powered devices and carbon emission related environmental cost are also important considerations.

For example, \citet{robinson2019large} trained a fully convolutional neural network (CNN) on over 55 terabytes of aerial imagery to create a high-resolution land cover map over the United States. Differences in seconds of running time per batch translate to hundreds of dollars in the cost of the final computation. Here, a larger, state-of-the-art model would incur a \textasciitilde270\% increase in the cost of the final computation for a fractional increase in performance metrics such as accuracy and intersection-over-union, and so a trade-off in favor of lowering the cost was made.

In wildlife conservation~\cite{weinstein2018computer} and accessibility applications~\cite{wolf2020democratizing}, models need to be deployed to edge or mobile devices of varying capacity. For instance, the Seeing AI\footnote{\url{https://www.microsoft.com/en-us/ai/seeing-ai}} mobile app, which helps people with vision impairment or low vision to better understand their surroundings, uses deep learning architectures specifically designed for low resource settings.

State-of-the-art deep learning models have large carbon footprints from training and operation~\cite{strubell2019energy}, which is a concern for AI4G projects in particular. Applications such as the one described in \citet{lacoste2019quantifying} allow the model developer to choose a data center location powered to a large extent by renewable energy sources and a cloud provider who offsets the remaining emissions.

\begin{center}
\fbox{
\begin{minipage}{\takeawaywidth em}
\textbf{Takeaway \textit{\takeaway}.} 
Carefully consider a project's constraints during deployment in advance before settling on a modeling approach.
\end{minipage}}
\end{center}

\subsection{Evaluation and metrics} \label{sec:modeling-eval-and-metrics}

Model validation is a crucial part of any AI project. In AI4G projects, validation metrics will not only need to measure how well the model is performing in standard ways (e.g. accuracy, AUC-ROC, intersection over union), but how well the model is performing with respect to any domain-specific requirements.

For example, the common part of commuters (CPC)~\cite{lenormand2012universal} is a domain-specific metric used in measuring how well predicted commuting flows align with ground truth data. This metric jointly considers all commuting flows together, as opposed to a common ML metric such as mean squared error that only considers pairwise errors between a single origin and destination. Reporting CPC reveals more about the overall structure of a predicted set of flows and is thus important to report in AI4G projects that consider commuter or migration flows~\cite{robinson2018machine}.

Another example comes from cancer imaging, where specialized extensions of the receiver operating characteristic (ROC) analysis are used to evaluate the lesion detection performance by radiology readers: localization ROC (LROC) quantifies not only the correct binary diagnosis, but also takes into account the accuracy of lesion localization within an image. The free-response operating characteristic (FROC) extends the notion of LROC to the multiple-lesion detection task~\cite{bandos2013subject}.

These and other domain-specific metrics can potentially be included in modeling as well as in evaluation. For example, if the domain-specific metric is differentiable with respect to the predicted quantity and computed on a per-sample basis, then it can be used in combination, or in place of, common loss functions when training models with gradient descent based methods. In the medical-imagery domain, such loss functions have been used to better capture domain-specific problem characteristics~\cite{taghanaki2019combo}.

In other cases, a domain-specific metric is not necessary, however domain experts will care little about commonly reported ML metrics. For example, mean average precision (mAP) involves averaging the precision of a model at all possible recall values. This average will include, for example, the precision of the model at 1\% recall which is not informative as such performance would never be acceptable. Precision@$k$, where $k$ is the lowest tolerable recall, is a more appropriate metric. As we discuss in Section 2, arriving at this metric involves extended dialogue with domain experts.

Finally, the data collection process by the PO can be imperfect, therefore model evaluation based solely on such datasets might be insufficient. For example, in the case of the anti-poaching PAWS project, the dataset is collected by limited park rangers via foot patrolling over a vast area. As such, many regions in the protected areas are not thoroughly covered by the rangers every month and the dataset does not perfectly represent the area under study. Building a long-term collaboration with POs to deploy machine learning solutions for pilot tests before a full commitment can bring important insights about the performance of the trained model in the wild~\cite{gholami2018adversary,xu2020stay}.

\begin{center}
\fbox{
\begin{minipage}{\takeawaywidth em}
\textbf{Takeaway \textit{\takeaway}.} Check if domain-specific metrics can be incorporated during training and validation of models and determine which ML metrics are relevant to solving the problem at hand.
\end{minipage}}
\end{center}

\subsection{Humans in the loop}

In the industry, continued data collection and user-supplied labels allow models to be improved over time, the so-called ``data flywheel'' effect~\cite{trautman2018virtuous,collins2019turning}.
In a similar manner, many scientific fields have come up with labeled datasets and models~\cite{vita2019immune,chanussot2020open,sun2020physical} that are continuously updated as labeling techniques (such as physical simulations or data acquisition methods) are improved. A common industry practice for improving model performance is to iterate on improving \textit{datasets} instead of iterating on improving \textit{models}~\cite{hohman2020understanding}. In both cases, having humans continually in the loop -- whether by labeling or tuning model behavior based on feedback from a deployed system -- provide large benefits to the overall project outcomes.

In general, the one-off nature of AI4G projects preclude this common way of improving model performance. Accumulating expert-annotated labels, even those created with efficiency gains enabled by the first version of the model, lies outside of formal infrastructures and therefore model re-training is not done as often.

At the same time, most of the models used in AI4G projects do not enable complete automation. For example, the output of a medical diagnosis model will be interpreted by health professionals, and final decision may be made based on the patient's clinical history and the presence of secondary signs and symptoms that are not captured by the model~\cite{miller2018artificial}. Here, a human \textit{must} be in the modeling loop evaluating every output of a model. More broadly, the output of \textit{all} AI4G models will be inspected by domain experts in the PO and their feedback will constitute a form of weak supervision that must be included in the modeling process in order to produce a suitable deliverable. For example, POs that rely on highly accurate land cover data will often need to make manual corrections to modeled outputs, and, as such, will have a difficult time using land cover predictions with artifacts such as rounded corners on building that are not easily correctable in GIS software. This type of feedback is only apparent after one iteration of modelling and inspecting the results with the PO. Glacier monitoring is another scenario where incorporating humans in the loop have been shown valuable. Baraka et. al proposed a glacier mapping tool that uses semantic segmentation predictions as a starting point and allows domain experts for easy adjustments of those predictions for a faster glacier mapping system~\cite{baraka2020machine}.
 
Thus, achieving a balance between having human feedback included in the modeling process and staying within-scope is a crucial part of finishing such AI4G projects. We have found active learning pipelines to be beneficial towards this end. With an active learning pipeline, participants at the PO can be engaged directly during the modeling process and will allow their feedback (in the form of labels) to be directly incorporated in the final deliverable.

We note that the active learning loop can also incorporate humans more tightly. For example, when humans are further allowed to choose the locations to label (versus being presented locations), and can observe the effect labeling those locations has on model output after a retraining period, they can more efficiently train land cover models~\cite{robinson2019human}. Finally, as active learning methods more tightly couple dataset collection with model training, they show promise in reducing the total amount of manual effort required to produce a final product. For instance, active learning training of camera trap species identification models has been found to match state-of-the-art accuracy with orders of magnitude fewer annotated training samples~\cite{sadegh2019deep}.

\begin{center}
\fbox{
\begin{minipage}{\takeawaywidth em}
\textbf{Takeaway \textit{\takeaway}.} AI4G projects require humans in the loop to some extent. Active learning pipelines can enable POs to engage with the modeling process directly during a project.
\end{minipage}}
\end{center}
\label{sec:modeling}

\section{Impact}
Lacking in the usual business indicators such as revenue and user engagement, one of the most difficult aspects of an AI4G project is measuring the degree to which it is successful, and weighing the success by the potential impact in advancing a PO's mission. In turn, an AI4G project's potential impact will not be realized without the PO or the broader communities adopting the technology. 
Indeed, there are many more press releases, blog posts, and promising published results than functioning AI systems actively ``doing good'' in the world. In this section, we attempt to understand why that is by exploring the three related issues of deployment, adoption and impact.

\subsection{Uphill path to deployment and adoption}
Unlike research developing novel techniques or theoretical understanding, AI4G projects necessitate the deployment of any developed models, and separately and often more difficult, the adoption of such technologies by the PO and related communities, for the effort to be meaningful~\cite{wagstaff2012machine}. This ``last mile'' problem can be especially challenging in AI4G projects since engineering is often not a focus for a PO or a research group.

In our experience, deployment entails three scenarios: \begin{enumerate}[label=\roman*)] 
\item a one-time scoring of relevant input data to produce derived data for the PO's downstream analysis and publication, 
\item a real-time API exposing the model for applications such as anti-poaching and invasive species monitoring, 
\item a batch processing mechanism triggered automatically or by the user to process a large quantity of raw data for recurring analysis, such as while processing conflict videos from a region for weapon detection~\cite{abdulrahim2021benetech}.\end{enumerate} The first scenario requires the least engineering effort beyond model development, but may not result in lasting impact. Real-time APIs have recurring cost, in addition to requiring upkeep and integration with the client application consuming model output. Batch processing could take advantage of discount cloud compute at low-traffic times and parallelize model scoring. It is often necessary to guide the PO in understanding whether they require real-time always-on model deployment or if offline batch processing is sufficient. There is also a lot of enthusiasm for deploying ML models on edge devices so that the PO can avoid uploading the raw data for processing in low-connectivity regions. In this case, it is important to communicate any trade-offs compressing the model through techniques such as quantization may have on performance~\cite{sheng2018quantization,wu2019machine}. To truly enable productivity gains from using AI tools and cloud infrastructure, the POs often need a much larger piece of software to orchestrate scoring raw data using the model and interact with the model outputs, of which Wildlife Insights is a notable example for accelerating wildlife surveys using computer vision models~\cite{ahumada2020wildlife}.

Adoption is the harder problem because, in many ways, it is outside of the control of the technical team. Identifying how ML metrics translate into time saved in the PO's workflows is paramount. Taking input from human-computer interaction experts may be helpful at this stage, as is thinking about how to integrate model output with the software used in downstream analysis. For example, being able to preview model outputs above a certain confidence threshold can help the domain expert to filter out input files that do not contain any subjects of interest; pre-populating the label field with the most common class may save many keystrokes during manual review~\cite{greenberg2020automated}. We have also found that open-sourcing the model development code builds trust in the model, and the code repository with discussion boards can act as a hub for the community involved in transfusing AI into their domain.

\begin{center}
\fbox{
\begin{minipage}{\takeawaywidth em}
\textbf{Takeaway \textit{\takeaway}.} Maintaining deployed models requires long-term engineering resource commitments. Focusing on time saved instead of pure ML metrics helps organizations adopt the technology.
\end{minipage}}
\end{center}

\subsection{Measuring impact}
Typical machine learning projects measure success in terms of model evaluation metrics (e.g. F1-score, ROC-AUC, etc.) (also see discussion in the last section) or business key performance indicator (KPIs) (e.g. click-through rate, daily active users, etc.). However, in the life cycle of AI4G projects, model evaluation metrics serve more as a basis of discussions with POs about a model's capabilities and limitations; the KPIs important to the POs are outcomes that can be several steps removed from the model outputs. It is important to learn about the PO's KPIs in the scoping phase of the project to inform the approach.

A challenge in creating \textit{lasting} impact comes from the lack of a business model for these AI4G endeavors. We are finding ways to step out of a funding mindset and grow the technical capabilities of the PO so that they could be self-sufficient in subsequent data collection and re-training efforts. This is another place where two-way communication is important (see Section 2): the technical team often does not get to see the impact of their work in the field. Maintaining a relationship with the PO after the technical portion of the project is complete to get updates on how their workflows have been impacted is important in maintaining a long-term collaboration and acts as part of a larger feed-back loop for solving the application problem.

In our engagements with POs, we do not concern ourselves with defining what is a positive impact against the final problem the PO aims to tackle. We rely on the domain experts at the PO to determine what the intended eventual impact is and to what extent a project furthers the PO's mission. More proximal to data modeling, in collaborating with POs on AI4G projects we have found two ways for AI techniques to realize impact: finding structure and insights from large datasets, and making domain experts' workflows more efficient so that they may scale out their work. Therefore, it may require working with the PO to find ways to track the immediate impact of an AI4G model on their data analysis or workflow efficiency, in addition to impacts on the PO's end mission.

\begin{center}
\fbox{
\begin{minipage}{\takeawaywidth em}
\textbf{Takeaway \textit{\takeaway}.} Domain experts within POs should define mission-related impacts. When quantification of direct model impact is needed, work with the PO to identify opportunities to quantify both immediate (workflow or analysis enhancement) and farther-removed (mission-related) impacts of the AI4G project.
\end{minipage}}
\end{center}
\label{sec:impact}

\section{Case studies}

\subsection{NLP to map Syrian conflict}

\noindent\textbf{Problem}:\\
The Carter Center (TCC) has been working on supporting a political solution to the wars in Syria\footnote{\url{https://www.cartercenter.org/countries/syria.html}}. Since 2012, TCC has initiated a conflict mapping project that analyzes an unprecedented volume of citizen-generated information about the conflict. Every week, TCC compiles a report using the information it receives from the Armed
Conflict Location and Event Data (ACLED) Project\footnote{\url{https://www.acleddata.com/data/}}~\cite{raleigh2010}, which curates news stories and articles recording incidents related to the war in Syria. This report is read by various committees in the UN, foreign ministries and NGOs. Given the weekly timeline (Takeaway 7), collating the incoming data into a structure suitable for their analysis manually has been difficult given the increasing volume of reports. Automating this curation process would reduce the thousands of hours of work needed by professional analysts.

\noindent\textbf{Solution}:\\
Our work automated this process by classifying the information into several categories such as shelling, artillery fire, and aerial bombardment. We helped TCC build a high-precision, neural network-based natural language processing (NLP) model that reclassifies the input conflict events at the granularity desired by TCC (Takeaway 6). This improvement in data processing allowed TCC employees to then focus on subsequent analysis of the conflict events (Takeaway 9).

\subsection{Mapping solar farms across India}

\noindent\textbf{Problem}:\\
At the end of 2020, India was only 2\% away from the target of 40\% installed non-fossil fuel electricity capacity, one of its Paris Climate Agreement targets~\cite{jaiswal2020climate}. While it is encouraging to see renewable energy production systems, such as solar farms, being rapidly built, it is also important to locate such installations in a way that avoids encroaching on the habitats of endangered species and other ecological reserves. An international conservation NGO has been working with states in India to create a tool for identifying areas where solar and wind developments are less likely to cause ecological harm. However, information on where solar installations are located is only available for two states, and so we worked together with the NGO to use satellite imagery to try identify solar installations across all of India. 

\noindent\textbf{Solution}:\\
Finding solar farms from satellite imagery is straight-forward to formulate as a semantic segmentation task, but the labels that were available for the project were both few and not in the format needed for this ML task: only 72 point labels of locations of solar farms in two states were available (Takeaway 3). To overcome this limitation, we first pre-trained a convolutional neural network to cluster pixels in the input satellite imagery by color (i.e. in an unsupervised manner). We then used an interactive training application to quickly fine-tune the network to segment the classes of interest and used this fine-tuned model to obtain weak segmentation labels for the entirety of the study area. These weak labels make it possible to train a supervised semantic segmentation network that was capable of accurately detecting solar farms. Solar farms found by this supervised  model were validated by analysts at the NGO. In total we were able to find and validate 1422 solar installations across India. The human-in-the-loop process we used was a crucial component in both training and evaluation, enabling ML models yield reliable results given the small amounts of labels available initially (Takeaway 9). Given the large area of interest, we must also rely on free satellite imagery, which is lower in resolution than commercial imagery; we accept this constraint to ensure our solution remains practically useful in the long term as the NGO updates the map each year (Takeaway 10). To reduce the number of false positive identifications, we incorporated OpenStreetMap data to remove areas of roads, snow and water bodies, a post-processing step informed by expertise in geospatial analysis (Takeaway 6).

\section{Conclusion}

Our work presents a broad overview of the considerations necessary while working on AI4G problems and the challenges encountered therein. We observe that the most useful AI4G projects result from working closely with specific stakeholders and understanding their operations and needs; attention to the particular characteristics of the problem while developing ML models, metrics and evaluation; a deep understanding of ethics and fairness concerns; a commitment to sound scientific and engineering practices and a transfer of technology that empowers the beneficiaries to understand and learn from the solution, and hopefully, adapt it with their changing needs. To support our observations we present several examples from our own experiences and relevant literature and summarize the learned lessons in takeaways. We hope that our endeavor helps researchers who are passionate about social good causes by bridging the gap between ML methodologies and their potential for relevant impact. However, we note that there are many problems and questions still outstanding, and that we are continually learning and growing our own repertoire in tackling challenging issues and working with POs. Becoming good at AI4G is a process that we are actively engaged in, and we hope others join us and learn with us.

\begin{acks}
We would like to thank Karthikeyan Ramamurthy for helpful feedback on an initial draft. We would also like to thank the many partner organizations who have collaborated with us over the past several years.
\end{acks}

\bibliographystyle{ACM-Reference-Format}
\balance
\bibliography{bibliography}

\end{document}